# Quantum Vacuum and Virtual Gravitational Dipoles: The solution to the Dark Energy Problem?

Dragan Slavkov Hajdukovic[1]
PH Division CERN
CH-1211 Geneva 23
dragan.hajdukovic@cern.ch
[1]On leave from Cetinje, Montenegro

**Abstract**
The cosmological constant problem is the principal obstacle in the attempt to interpret dark energy as the quantum vacuum energy. We suggest that the obstacle can be removed, i.e. that the cosmological constant problem can be resolved by assuming that the virtual particles and antiparticles in the quantum vacuum have the gravitational charge of the opposite sign. The corresponding estimates of the cosmological constant, dark energy density and the equation of state for dark energy are in the intriguing agreement with the observed values in the present day Universe. However, our approach and the Standard Cosmology lead to very different predictions for the future of the Universe; the exponential growth of the scale factor, predicted by the Standard Cosmology, is suppressed in our model.

## 1. Introduction

The nature of dark energy, invoked to explain the accelerated expansion of the Universe, is a major mystery in the theoretical physics and cosmology. From the purely mathematical point of view, adding a positive cosmological constant term to the right-hand side of the Einstein equation, can account for the observed accelerated expansion. However no one knows what is the physics behind such an ad hoc introduction of the cosmological constant. In principle, the cosmological constant $\Lambda$, may be interpreted as a cosmological fluid with a constant density ($\rho_{de}$) and negative pressure ($p_{de} = -\rho_{de} c^2$), i.e. $\Lambda = 8\pi G \rho_{de}/c^2$, but the physical nature of such a hypothetical fluid stays unknown.

The most elegant and natural solution would be to identify dark energy with the energy of the quantum vacuum predicted by the Quantum Field Theory (QFT); but the trouble is that QFT (for a classical Review see Weinberg, 1989) predicts the energy density of the vacuum to be many orders of magnitude greater than the observed dark energy density and the corresponding cosmological constant:

$$\rho_{de} \approx 7.2 \times 10^{-27} \, kg/m^3 \qquad (1)$$

$$\Lambda \approx 1.4 \times 10^{-52} \, m^{-2} \qquad (2)$$

Summing the zero-point energies of all normal modes of some field of mass $m$ up to a wave number cut-off $K_c \gg m$, QFT (Weinberg, 1989) yields a vacuum energy density (with $\hbar = c = 1$)

$$\langle \rho_{ve} \rangle = \int_0^{K_c} \frac{k^2 \sqrt{k^2 + m^2}}{(2\pi)^2} dk \approx \frac{K_c^4}{16\pi^2} \qquad (3)$$

or reintroducing $\hbar$ and $c$ and using the corresponding mass cut-off $M_c$ instead of $K_c$:

$$\rho_{ve} = \frac{1}{16\pi^2} \left(\frac{c}{\hbar}\right)^3 M_c^4 \equiv \frac{\pi}{2} \frac{M_c}{\lambda_{Mc}^3} \qquad (4)$$

where $\lambda_{Mc}$ denotes the (non-reduced) Compton wavelength corresponding to $M_c$. If we take the Planck scale (i.e. the Planck mass) as a cut-off, the vacuum energy density calculated from Eq. (4) is $10^{121}$ times larger than the observed dark energy density (1). If we only worry about zero-point energies in quantum chromodynamics, Eq. (4) is still $10^{41}$ times larger than the Eq. (1). Even if the Compton wavelength of an electron is taken as cut-off, the result exceeds the observed value by nearly 30 orders of magnitude. This huge discrepancy is known as the cosmological constant problem (See Weinberg, 1989 for more details).

In the present Letter, by assuming that antiparticles have a negative gravitational charge (i.e. that the virtual particle-antiparticle pairs in the quantum vacuum may be considered as gravitational dipoles) we show how dark energy density and the cosmological constant can be accurately estimated and understood as a signature of the quantum vacuum. These estimates lead to a new equation of state for dark energy and in major changes in the prediction of the cosmological field equations. In particular we point out that the exponential growth of the cosmological scale factor (predicted by Standard Cosmology) must be suppressed.

## 2. Gravitational charge and energy of the quantum vacuum

Let us assume that the gravitational charge $m_g$ of a particle, and the gravitational charge $\bar{m}_g$ of an antiparticle have the opposite sign, i.e. that the relation

$$m_g + \bar{m}_g = 0 \tag{5}$$

is valid for all particle-antiparticle pairs. The idea that antimatter could have a negative gravitational charge is older than half a century (See Nieto, 1991 for a Review) and reinforced by the recent theoretical arguments (Villata, 2011). However, before our work, the hypothesis of the gravitational repulsion between matter and antimatter has never been combined with the well established fact of the existence of the quantum vacuum. There are two consequences of the hypothesis (5):

(a) The gravitational charge of the quantum vacuum is zero (in the same way as the electric charge is zero). Because of this cancelation of the gravitational charges, the quantum vacuum, if not perturbed by the long-living baryonic matter, naturally has the cosmological constant equal to zero.

(b) A virtual particle-antiparticle pair in the quantum vacuum, may be considered as a gravitational dipole with the gravitational dipole moment

$$\vec{p} = m\vec{d}\,;\ |\vec{p}| < \frac{\hbar}{c} \tag{6}$$

Here, by definition, the vector $\vec{d}$ is directed from the antiparticle to the particle, and has a magnitude equal to the distance between them. The inequality in (6) follows from the fact that the size of the virtual pair must be smaller than the reduced Compton wavelength $\lambdabar_m = \hbar/mc$ (for larger separation a virtual pair becomes real). Hence, $|\vec{p}|$ should be a fraction of $\hbar/c$. In the quantum field theory, the volume occupied by a virtual pair is $\lambda_m^3$ (where $\lambda_m \equiv 2\pi\hbar/mc$ is the non-reduced Compton wavelength). As argued in previous papers (Hajdukovic 2010a, 2010b, 2011a, 2012) the pions (as the simplest quark-antiquark pairs) probably dominate the quantum vacuum and $\lambda_m$ should be identified with the Compton wavelength $\lambda_\pi$ of a pion. Hence, the number of the virtual gravitational dipoles per unit volume has a constant value

$$N_0 \propto \frac{1}{\lambda_\pi^3} \tag{7}$$

Recently (Hajdukovic, 2011a, 2012) it was argued that the phenomena attributed to the hypothetical dark matter, can be explained by the gravitational polarization of the quantum vacuum in the gravitational field of the baryonic matter of the Universe. In the present Letter we argue that the phenomena, for which dark energy has been invoked, can also be explained in the framework of the quantum vacuum enriched with the gravitational repulsion between matter and antimatter. More precisely, we suggest that what we call dark energy is, in fact, the energy of the virtual gravitational dipoles inhabiting the quantum vacuum. This energy of the gravitational dipoles in a gravitational field can be estimated in two different ways, as described below.

Firstly, let us remember that the energy of an electric dipole $\vec{p}$ in the electric field $\vec{E}$ is determined by the scalar product $\vec{p} \cdot \vec{E}$. In the case of a gravitational dipole in the gravitational field $\vec{g}$ the energy of the dipole would be determined by $\vec{p} \cdot \vec{g}$. Consequently, according to equations (6) and (7), the order of magnitude for the energy density and the gravitational charge density is respectively:

$$\rho_{ve} \propto \frac{1}{\lambda_\pi^3} \frac{\hbar}{c} g; \quad \rho_v \propto \frac{1}{\lambda_\pi^3} \frac{\hbar}{c^3} g \tag{8}$$

In the case of the expansion of the Universe it is natural to take $g = \ddot{R}$, where $R$ is the cosmological scale factor (defined in (12)) and $\ddot{R}$ is the acceleration of the expansion, which according to observations should have the present day value $\ddot{R}_0 \approx 5.5 \times 10^{-9} \, m/s^2$. What a surprise. With $g = \ddot{R}_0$, the second of relations (8) leads to the numerical result that is only about 4 times greater than the observed values (1) and (2). It is very intriguing that such a simple reasoning gives the right order of magnitude for dark energy density and the cosmological constant.

For the purpose of the forthcoming considerations, let us write the second of proportionalities (8) in the form of equality:

$$\rho_v = \frac{C}{2\pi} \frac{1}{\lambda_\pi^3} \frac{\hbar}{c^3} \ddot{R} \tag{9}$$

where (in order to fit the observed value (1)), the best choice for the constant $C$ is $C \approx 3/2$. The factor $2\pi$ is introduced for the easier comparison with the famous Hawking-Unruh temperature (Hawking, 1975; Unruh. 1976) for an accelerated observer in the vacuum.

$$k_B T = \frac{1}{2\pi} \frac{\hbar}{c} g \tag{10}$$

Our result (9) appears to be an unexpected generalisation of the Hawking-Unruh temperature (10).

It is amusing that dark energy density (1) can be accurately estimated in a different way, taking as the starting point, the "wrong" result (4) derived by quantum field theory. In fact, the result (4) is not wrong, but instead of the gravitational charge density, it determines the inertial mass density of the quantum vacuum. Hence, the essence of what we call the cosmological constant problem may be that it is the exact calculation of the wrong (i.e., non-relevant) quantity.

Secondly, the result (4) is a consequence of the assumption that inertial and gravitational mass are equivalent; the source of gravitation are always gravitational monopoles with a positive gravitational mass. But what if the sources of gravitation are gravitational dipoles? Let us compare the gravitational field produced by two positive monopoles (at mutual distance $\lambda_\pi$) and a dipole (i.e. a positive and a negative monopole also at distance $\lambda_\pi$). In full analogy with the electric dipole, the gravitational field produced by a dipole at large distance $R \gg \lambda_\pi$ is $\lambda_\pi/R$ weaker than the corresponding field produced by two monopoles of the same sign. Hence, the result (4) should be modified according to a simple rule: use the mass of a pion as a cut-off on the right-hand side of Eq. (4) and multiply it by $\lambda_\pi/R$. So, using (4) as the starting point leads to the following estimate of the order of magnitude of the gravitational charge density

$$\rho_v \propto \frac{m_\pi}{\lambda_\pi^2} \frac{1}{R} \qquad (11)$$

Using the value $R = R_0 \approx 1.65 \times 10^{27} m$ (i.e. the present day value of the scale factor, leads once again to the result only about 4 times greater than (1) and (2).

## 3. A new equation of state for dark energy

In order to understand the impact of relation (9) let us remember the basic points of the Standard Cosmology.

As well known, the cosmological principle (i.e. the statement that at any particular time the Universe is isotropic about every point) determines the Friedman-Robertson-Walker metric

$$ds^2 = c^2 dt^2 - R^2(t)\left[\frac{dr^2}{1-kr^2} + r^2\left(d\theta^2 + \sin^2\theta d\vartheta^2\right)\right] \qquad (12)$$

where $k = +1;\ k = -1;\ k = 0$ correspond respectively to closed, open and flat Universe.

The dynamics of the above space-time geometry is entirely characterised by the scale factor $R(t)$. In order to determine the function $R(t)$, the Einstein equation $G_{\mu\nu} = -(8\pi G/c^4)T_{\mu\nu}$ must be solved. While the Einstein tensor $G_{\mu\nu}$ is determined by metric (12); we need a model for the energy-momentum tensor ($T_{\mu\nu}$) of the content of the Universe. In view of homogeneity and isotropy of the Universe, a reasonable approximation is to assume the energy-momentum tensor of a perfect fluid; characterised at each point by its proper density $\rho$ and the pressure $p$ in the instantaneous rest frame. Assuming that the cosmological fluid in fact consists of several distinct components (for example, matter, radiation and the vacuum) the final results are cosmological field equations, which may be written in the form:

$$\ddot{R} = -\frac{4\pi G}{3}R\sum_n\left(\rho_n + \frac{3p_n}{c^2}\right) \qquad (13)$$

$$\dot{R}^2 = \frac{8\pi G}{3}R^2\sum_n\rho_n - kc^2 \qquad (14)$$

However, it is still not enough. In order to solve cosmological field Equations (13) and (14), we need equation of state for every component of the cosmological fluid. The most used equation of state, relating pressure and energy density is:

$$p_n = w_n \rho_n c^2 \qquad (15)$$

where the equation-of state-parameter $w_n$ is a constant.

The density $\rho_n$ of a fluid, satisfying the above equation of state, transforms according to the power-law

$$\rho_n = \rho_{n0}\left(\frac{R_0}{R}\right)^n \qquad (16)$$

where, as usually, index 0 denotes the present day value.

Radiation, matter and dark energy (when identified with cosmological constant) are modelled respectively with $w_r = 1/3$ (i.e. $n = 4$); $w_m = 0$ (i.e. $n = 3$) and $w_\Lambda = -1$ (i.e. $n = 0$). In fact, matter is attributed a non-zero density $\rho_m$ and zero pressure $p_m = 0$; that is why we sometimes refer to it as pressureless matter. Dark energy is characterized with $w_\Lambda = -1$, i.e., a constant energy density $\rho_\Lambda c^2$ and a constant negative pressure $p_\Lambda = -\rho_\Lambda c^2$. In the standard $\Lambda CDM$ cosmology, the dark energy term provides a continual acceleration to the Universe.

Now, let us consider how this picture is changed by the inclusion of a simple matter density $\rho_m$ and a dark energy term of the form of equation (9). It is easy to check that simultaneous validity of equations (9) and (13) is possible only if dark energy has the following equation of state:

$$p_v = -\frac{1}{3}\left[\frac{\rho_m}{\rho_v} + 1 + \frac{3}{2C}\frac{\lambda_\pi^3}{L_P^2}\frac{1}{R}\right]\rho_v c^2 \equiv w_{eff}\rho_v c^2 \qquad (17)$$

where $L_P \equiv \sqrt{\hbar G/c^3}$ is the Planck length. Hence, the old equation of state (15) should be replaced by the new one (17); with the inevitable major consequences in Cosmology, The equation (17) deserves the following comment: while the equations (15) and (17) are very different, they may be considered as identical in the present day Universe. In fact, using the best known numerical values (Nakamura et al., 2010) for $\rho_{m0}$, $\rho_{v0}$ and $R_0$, the Eq.(17) leads to $p_{v0} \approx -0.99\rho_{v0}c^2$ (i.e. $w_{eff} \approx -0.99$) which is indistinguishable from the standard choice $w_\Lambda = -1$ in Eq.(15). However for large $R$ i.e. $R \gg R_0$, $p_v$ approaches the maximum $-\rho_v c^2/3$ (i.e. $w_{eff} = -1/3$); hence the future of the Universe should be quite different from the standard $\Lambda CDM$ cosmology.

*4. Evolution of the dark energy density*

In order to find how vacuum energy density evolves with the cosmological scale factor $R(t)$ it is appropriate to start with the second cosmological equation (14). The first step is to differentiate (14) with respect to time and after differentiation to introduce $\ddot{R}$ (determined by (13)) and $\rho_m$ and $\dot{\rho}_m$ determined by (16) with $n = 3$. The result is the differential equation:

$$R^2 d\rho_v + 2\rho_v R dR - a_0\rho_v R_0 dR = \rho_{m0} R_0\left(\frac{R_0}{R}\right)^2 dR; \quad a_0 \equiv \frac{3}{2C}\frac{\lambda_\pi^3}{L_P^2 R_0} \approx 1.62 \qquad (18)$$

having as solution

$$\rho_v = \rho_{v0}\left(\frac{R_0}{R}\right)^2\left\{\left(1 + \frac{1}{a_0}\frac{\rho_{m0}}{\rho_{v0}}\right)e^{a_0\left(1-\frac{R_0}{R}\right)} - \frac{1}{a_0}\frac{\rho_{m0}}{\rho_{v0}}\right\} \qquad (19)$$

Now, equations (9) and (19) give

$$\ddot{R} = \frac{a_0\Omega_{v0}}{2}R_0 H_0^2\left(\frac{R_0}{R}\right)^2\left\{\left(1 + \frac{1}{a_0}\frac{\Omega_{m0}}{\Omega_{v0}}\right)e^{a_0\left(1-\frac{R_0}{R}\right)} - \frac{1}{a_0}\frac{\Omega_{m0}}{\Omega_{v0}}\right\} \qquad (20)$$

Let's note that in the last equation I have introduced usual dimensionless density parameters $\Omega_m$ and $\Omega_v$, instead of $\rho_m$ and $\rho_v$.

The equation (20) shows that the scale factor has a critical value $R_{crit}$, determined by

$$\frac{R_0}{R_{crit}} = 1 + \frac{1}{a_0}\ln\frac{a_0\Omega_{v0} + \Omega_{m0}}{\Omega_{m0}} \approx 2 \qquad (21)$$

so that $\ddot{R} < 0$ when $R < R_{crit}$ and $\ddot{R} > 0$ when $R > R_{crit}$. Hence the accelerated expansion of the Universe has started when Universe was about half of its present size (i.e. $R_{crit} \approx 0.50 R_0$); which is slightly earlier than the prediction of the standard $\Lambda CDM$ cosmology: $R_{crit} = (\Omega_{m0}/2\Omega_{v0})^{1/3} R_0 \approx 0.56 R_0$.

However, for the late time Universe (i.e. when $R \gg R_0$) there is a dramatic difference between acceleration (20) *decreasing* as $(R_0/R)^2$ and the result of standard cosmology predicting an acceleration $\ddot{R}_{sc}$ *increasing* linearly with the scale factor i.e. $\ddot{R}_{sc} = \Omega_{v0} H_0^2 R$. Consequently, in the late-time Universe, cosmological scale factor is a linear function of time, contrary to the prediction of the standard ΛCDM cosmology which predicts a scale factor increasing exponentially with time.

## 5. *Revision of Dirac's relation for mass of a pion*

A simple transformation of the equation (9) gives

$$m_\pi^3 = \frac{\hbar^2}{cG} H \left\{ 3\pi^3 c \frac{H\Omega_v}{\ddot{R}} \right\} \qquad (22)$$

The incomplete relation (22), without the term in brackets, i.e. proportionality

$$m_\pi^3 = \frac{\hbar^2}{cG} H \qquad (23)$$

was known to Dirac (Dirac,1937) and Weinberg (Weinberg, 1972), but there are problems with relation (23). The Hubble parameter (and hence the right-hand side of relation (23)) is a function of the age of the Universe; while the left-hand side of the same relation should be a constant. In fact, in order to get the right mass of a pion, we are forced to choose $H = H_0$ in relation (23), and even so the left-hand side is about one order of magnitude greater than the right one. In order to save relation (23) as a fundamental one, Dirac has suggested that the ratio $H/R$ must stay constant with time; hence introducing a varying gravitational "constant" not supported by observations (Weinberg, 1972). The alternative with constant ratio $H/c$, introducing a varying speed of light was considered as well (Alfonso-Faus, 2008). My position is quite different: relation (23) is considered as an incomplete relation which must be completed in an appropriate way. In fact, without invoking varying "constants" the conjecture (9) leads to the "missing" dimensionless term in brackets, having needed numerical value close to 12, and assuring that the right hand side of (22) does not change with the expansion of the Universe.

Of course, (22) corresponds only to the conjectured equation of state (17). The corresponding modification of relation (23) for the standard equation of state is given in Hajdukovic, 2010a and 2010b.

## 6. *Comments*

For a physicist "spoiled" with the well developed theories (Special Relativity, General Relativity, Quantum Field Theory, Classical Electrodynamics...), the standard ΛCDM cosmology cannot yet be called a theory. It is rather a pragmatic platform designed to fill the gap between observation and cosmological theory; it leaves unexplained the nature of both dark matter and dark energy. We are probably in the throes of a new scientific "revolution", which would replace ΛCDM cosmology by a satisfactory physical quantum gravitational theory. The present Letter belongs to a wave of new publications (for instance: Alfonso-Faus 2011, Benoit-Levy 2011, Fahr 2012, Fullana 2012, Gine 2012, Santos 2010 and 2011, Villata 2011 and 2012) rich in ideas which may be or may not be useful (we never know in advance) on the difficult way towards the future complete theory. In the useful crowd of new ideas (if one of many ideas is the right one it is the scientific victory of a generation) my work can easily be distinguished: for the first time the reality of the quantum vacuum is combined with the hypothesis that gravitation can be repulsive. In the case of the gravitational repulsion between matter and antimatter, quantum vacuum can be considered as a fluid of gravitational dipoles; a hypothesis very rich in testable consequences (Hajdukovic 2011a, 2011b, 2011c, 2011d, 2012).

Concerning the present Letter, if the relation (9) is correct, the pressure of the perfect fluid modelling dark energy is function of three variables: dark energy density, matter density and the size of the Universe. It should not be a surprise. The standard equation of state (15) assuming that pressure depends only on the dark energy density is presumably an oversimplification neglecting a possible impact of baryonic matter on the physical vacuum. It seems plausible to me that matter acts as an external gravitational field "inducing" a certain pressure in the physical vacuum. Hence pressure should have two components: "induced" pressure and "internal" pressure; what is incorporated in the new equation of state (17).

Let us underline once again the conclusion from the end of section 4, that the exponential growth of the scale factor, predicted by the Standard Cosmology, is suppressed in our model. It is important to note that simultaneously with our work, on the basis of thermodynamic considerations (Fahr, 2012), it was suggested that dark energy density "drops off with the expansion inversely proportional to the square of the cosmic scale", what is support to our results (19) and (20) which are basis for conclusion that the exponential expansion is suppressed.

In conclusion, after a series of intriguing arguments (Hajdukovic 2011a, 2012) that dark matter could be explained in the framework of the quantum vacuum enriched with the gravitational repulsion between matter and antimatter, we have presented the first indications that the dark energy may be understood within the same framework. Additionally, the effects related to the gravitational version of the Schwinger mechanism (i.e. the conversion of a virtual pair into a real one by the strong gravitational field) have potential to explain why we live in the Universe dominated with matter and to eliminate the need for inflation in cosmology (Hajdukovic 2011b), but they also lead to a radically different picture of black holes (Hajdukovic 2011c) or theoretical prediction of mass of neutrino (Hajdukovic 2011d). It completes the first, embryonic phase of our research; the more detailed studies are in progress.